\documentclass[12pt]{iopart}
\begin{document}
\title[Traveling wave solutions and steady-state of PASEP]{Discrete-time analysis of traveling wave solutions and steady-state of PASEP with open boundaries}
\author{S. R. Masharian$^1$, F. Zamani$^2$}
\address{$^1$ Islamic Azad University, Hamedan Branch, Hamedan, Iran}
\address{$^2$ Bu-Ali Sina University, Physics department, Hamedan, Iran}
\ead{masharian@iauh.ac.ir}
\begin{abstract}
We consider the dynamics of a single shock in a Partially Asymmetric Simple Exclusion Process (PASEP) on a finite lattice with
open boundaries in sublattice-parallel updating scheme. We then construct the steady-state of the system by considering
a linear superposition of these shocks. It is shown that this steady-state can be also written in terms of a product
of four non-commuting matrices. One of the main results obtained here is that these matrices have exactly the same generic structure of
the matrices first introduced in \cite{JM09} indicating that the steady-state of a one-dimensional driven-diffusive system can be
written as a linear superposition of product shock measures. It is easy now to explain the two-dimensional
matrix representation of the PASEP with parallel dynamics introduced in \cite{ER96,HP97}.
\end{abstract}
\maketitle
Some of the one-dimensional driven-diffusive systems with open boundaries have, among several unique properties, the feature that the dynamics of
a shock distribution in the system is similar to that of a random walker provided that some constraints on the reaction parameters are fulfilled. For instance, it is known that in the Partially Asymmetric Simple Exclusion Process (PASEP) with open boundaries multiple shocks might appear in the density profile of the particles \cite{BS02}. The PASEP usually refers to a driven-diffusive system of classical particles with hard-core interactions which perform biased random walk on a discrete lattice of finite length with open boundaries. When a boundary is open, the particles are allowed to enter or leave the lattice from there. The shocks in the PASEP, which are defined as the sharp discontinuities in the density profile of the particles, might appear in both the continuous- and discrete-time updating schemes \cite{BS02,PS00,KJS03}. \\
The steady-state of the PASEP can be easily calculated when a product shock measure in the system has a simple random walk dynamics. In this case the steady-state can be written as a linear superposition of the shocks in different positions on the lattice. It has been shown that the front of the product shock measure in the PASEP (single shock in this case), has a simple random walk dynamics provided that there are some constraints on the reaction rates (probabilities) in continuous-time (discrete-time) update. The diffusion and scattering of shocks in the PASEP on an infinite lattice and in continuous-time update has been studied in \cite{BS02}. The dynamics of a shock distribution as an initial state for the PASEP in discrete-time update has also been investigated in \cite{PS00}.\\
In this paper we consider the PASEP with open boundaries in a discrete-time update. We show that the shock will reflect from the boundaries of the lattice if we consider a finite size lattice. We calculate the reflection probabilities explicitly. In this case the equations of motion of the shock front are closed and one can explicitly construct the steady-state of the PASEP in terms of a linear superposition of these shocks. On the other hand, since the steady-state of the system is unique, we can find the steady-state of the system using a matrix product method \cite{DEHP93} which has been reviewed in \cite{BE07}. It has been claimed that when the steady-state of a system with nearest-neighbor interactions and open boundaries can be expressed in terms of a linear superposition of the product shock measures with a single shock front and random walk dynamics, the quadratic algebra of the system has a two-dimensional matrix representation \cite{JM07}. This matrix representation has a generic structure and matrix entries can be written in terms of the shock front hopping rates and the densities of the particles on the sides of the shock front. It seems that the same idea can be used in discrete-time update \cite{JM09}. After finding the equations of motions for a single product shock measure in the PASEP with open boundaries on a finite lattice in the sublattice-parallel updating scheme, we calculate the reflection probabilities of the shock front from the boundaries. Considering a linear superposition of these shocks we construct the steady-state of the system. Finally we show that if the steady-state of the system is calculated using the matrix product method, then a two-dimensional matrix representation of the form first introduced in \cite{JM09} will be sufficient. This helps us explain the two-dimensional matrix-representation of the quadratic algebra of the PASEP first introduced in \cite{ER96,HP97} in terms of the shock characteristics of the system.\\
We start with the equations of motion of a single shock in the PASEP in sublattice-parallel dynamics. We consider a discrete lattice of length $2L$. Each lattice site can be occupied by at most one particle or it is empty. In the bulk of the system each particle hops to the left (right) neighboring site with the probability $q$ ($p$) provided that it is not already occupied. The particles can enter into the system from the leftmost (rightmost) lattice site with the probability $\alpha$ ($\delta$). The particles can also leave the system from the leftmost (rightmost) lattice site with the probability $\gamma$ ($\beta$). The discrete-time evolution of the probability distribution is governed by the following master equation:
\begin{equation}
\label{time evol}
T \vert P(t) \rangle
= \vert P(t+1) \rangle.
\end{equation}
in which $T$ is the transfer matrix which is defined as follows: in the sublattice-parallel dynamics, the bulk dynamics consists of two half time steps.
In the first half time step even lattice sites i.e. the pairs of neighboring sites ($2k,2k+1$) for $k=1,\cdots,L-1$ and also the first and the last lattice sites are updated. From the first and the last lattice sites the particles can be injected or extracted with the above mentioned probabilities. In the second half time step only the odd lattice sites i.e. the pairs of neighboring sites ($2k-1,2k$) for $k=1,\cdots,L$ are updated; therefore, the transfer matrix $T$ is given by the multiplication of two factors $T=T_2 T_1$ defined as:
\begin{eqnarray}
\label{T1}
T_1 &=& {\cal L} \otimes {\cal T} \otimes \ldots \otimes
        {\cal T} \otimes {\cal R} \,\;=\;\,
{\cal L} \otimes {\cal T}^{\otimes (L-1)} \otimes {\cal R} \nonumber\\
\label{T2}
T_2 &=& \hspace{3.8mm} {\cal T} \otimes {\cal T} \otimes \ldots \otimes
        {\cal T} \hspace{3.8mm} \,\;=\;\,  {\cal T}^{\otimes L} \nonumber
\end{eqnarray}
where ${\cal T}$, ${\cal L}$ and ${\cal R}$ are given by:
\begin{equation}
\label{trans mat}
\fl
{\cal T} \;=\; \left(
\begin{array}{cccc}
1 & 0 & 0 & 0 \\
0 & 1-q & p & 0 \\
0 & q & 1-p & 0 \\
0 & 0 & 0 & 1
\end{array} \right),\;
{\cal L} \;=\; \left(
\begin{array}{cc}
1-\alpha & \gamma \\
\alpha & 1-\gamma
\end{array} \right),\;
{\cal R} \;=\; \left(
\begin{array}{cc}
1-\delta & \beta \\
\delta & 1-\beta
\end{array} \right).
\end{equation}
The matrix $\cal T$ is written in the basis $(00,01,10,11)$ when $0$ stands for an empty lattice site and $1$ stands for an occupied lattice site. The matrices $\cal L$ and $\cal R$ are also written in the basis $(0,1)$. \\
Following \cite{JM09} let us define two different product shock measures. We denote the shocks at even sites $2k$ ($k=1,\cdots,L$) as $\vert \mu_{2k}\rangle$ and at odd sites $2k+1$ ($k=0,\cdots,L$) as $\vert \mu_{2k+1} \rangle$ respectively:
\begin{equation}
\label{shocks}
\fl
\begin{array}{l}
\vert \mu_{2k} \rangle=
\left(\begin{array}{c}
1-\rho_{1}^{o} \\ \rho_{1}^{o}
\end{array}\right)\otimes
\left(\begin{array}{c}
1-\rho_{1}^{e} \\ \rho_{1}^{e}
\end{array}\right)\otimes \cdots \otimes
\underbrace{ \left(\begin{array}{c}
1-\rho_{2}^{e} \\ \rho_{2}^{e}
\end{array}\right)}_{2k}\otimes
\left(\begin{array}{c}
1-\rho_{2}^{o} \\ \rho_{2}^{o}
\end{array}\right)\otimes \cdots \otimes
\underbrace{\left(\begin{array}{c}
1-\rho_{2}^{e} \\ \rho_{2}^{e}
\end{array}\right)}_{2L},
\\
\vert \mu_{2k+1} \rangle=
\left(\begin{array}{c}
1-\rho_{1}^{o} \\ \rho_{1}^{o}
\end{array}\right)\otimes
\left(\begin{array}{c}
1-\rho_{1}^{e} \\ \rho_{1}^{e}
\end{array}\right)\otimes \cdots \otimes
\underbrace{ \left(\begin{array}{c}
1-\rho_{2}^{o} \\ \rho_{2}^{o}
\end{array}\right)}_{2k+1}\otimes
\left(\begin{array}{c}
1-\rho_{2}^{o} \\ \rho_{2}^{o}
\end{array}\right)\otimes\cdots \otimes
\underbrace{\left(\begin{array}{c}
1-\rho_{2}^{e} \\ \rho_{2}^{e}
\end{array}\right)}_{2L}
\end{array}
\end{equation}
in which $\rho_{i}^{o}$ and $\rho_{i}^{e}$ ($i=1,2$) stand for the density of particles at odd and even lattice sites, respectively.\\
As can be seen for $\vert \mu_{2k} \rangle$, the shock front lies between the lattice sites $2k-1$ and $2k$ while for $\vert \mu_{2k+1}\rangle$ the shock front lies between the lattice sites $2k$ and $2k+1$. For the mathematical consistency, we have defined an auxiliary lattice site $2L+1$; therefore, the shock $\vert \mu_{2L+1}\rangle$ indicates a flat distribution of particles with densities $\rho_{1}^{o}$ and $\rho_{1}^{e}$ at odd and even lattice sites respectively. In this case the shock front can be considered to be between the lattice sites $2L$ and $2L+1$. Note that the equations of motion for a single shock on an infinite lattice without boundaries has already been found in \cite{PS00}.\\
Using (\ref{time evol}) and (\ref{trans mat}) it can easily be verified that the shocks in (\ref{shocks}) evolve in time according to the following equations:
\begin{equation}
\label{eqs}
\fl
\begin{array}{l}
T\vert \mu_{2k} \rangle=\delta_{l} \vert \mu_{2k-1} \rangle+\delta_{r} \vert \mu_{2k+1} \rangle+\delta_{s} \vert \mu_{2k} \rangle \;\; 1 \leq k \leq L,\\ \\
T\vert \mu_{2k+1} \rangle=\delta_{l} \delta_{s} \vert \mu_{2k} \rangle +\delta_{r} \delta_{s} \vert \mu_{2k+2} \rangle + \delta_{r}^{2} \vert \mu_{2k+3} \rangle+\delta_{l}^2\vert \mu_{2k-1} \rangle+(\delta_{s}+2\delta_{l}\delta_{r})\vert \mu_{2k+1} \rangle\;\; 1 \leq k \leq L-1,\\ \\
T\vert \mu_{1} \rangle=\overline{\delta}_{r} \overline{\delta}_{s} \vert \mu_{2} \rangle+\overline{\delta}_{r}^2 \vert \mu_{3} \rangle+(1-\overline{\delta}_{r}\overline{\delta}_{s}-\overline{\delta}_{r}^2) \vert \mu_{1} \rangle, \\ \\
T\vert \mu_{2L+1} \rangle=\overline{\overline{\delta}}_{l} \overline{\overline{\delta}}_{s} \vert \mu_{2L} \rangle+\overline{\overline{\delta}}_{l}^2 \vert \mu_{2L-1} \rangle+(1-\overline{\overline{\delta}}_{l}\overline{\overline{\delta}}_{s}-\overline{\overline{\delta}}_{l}^2) \vert \mu_{2L+1} \rangle
\end{array}
\end{equation}
provided that we define the densities and the shock front hopping probabilities as:
\begin{equation}
\label{defs1}
\fl
\begin{array}{l}
\rho_1^{o}=\frac{1-p}{1-p+k_{+}(\alpha,\gamma)} \; , \; \rho_2^{o}=\frac{k_{+}(\beta,\delta)}{1-q+k_{+}(\beta,\delta)} \; , \;
\rho_1^{e}=\frac{1-q}{1-q+k_{+}(\alpha,\gamma)} \; , \; \rho_2^{e}=\frac{k_{+}(\beta,\delta)}{1-p+k_{+}(\beta,\delta)}, \\ \\
\delta_{r}=\frac{p(1-q)+qk_{+}(\beta,\delta)}{1-q+k_{+}(\beta,\delta)} \; , \; \delta_{l}=\frac{p(1-q)+qk_{+}(\alpha,\gamma)}{1-q+k_{+}(\alpha,\gamma)} \; , \; \delta_s=1-\delta_r-\delta_l, \\ \\
\overline{\delta}_{r}=\sqrt{\frac{(\rho_2^{e}-\rho_2^{o})(-\alpha+(\alpha-1+\gamma)\rho_2^{o}+\rho_2^{e})}{(\rho_2^{e}-\rho_1^{o})(\rho_2^{e}-\rho_1^{e})}}\; , \; \overline{\delta}_{l}=1-\sqrt{\frac{(\rho_2^{e}-\rho_1^{e})(-\alpha+(\alpha-1+\gamma)\rho_2^{o}+\rho_2^{e})}{(\rho_2^{e}-\rho_1^{o})(\rho_2^{e}-\rho_2^{o})}}  \; , \;
\overline{\delta}_s=1-\overline{\delta}_r-\overline{\delta}_l, \\ \\
\overline{\overline{\delta}}_{r}=1-\sqrt{\frac{(\rho_1^{o}-\rho_2^{o})(-\delta+(\delta-1+\beta)\rho_1^{e}+\rho_1^{o})}
{(\rho_2^{e}-\rho_1^{o})(\rho_1^{e}-\rho_1^{o})}}\; , \;
\overline{\overline{\delta}}_{l}=\sqrt{\frac{(\rho_1^{e}-\rho_1^{o})(-\delta+(\delta-1+\beta)\rho_1^{e}+\rho_1^{o})}
{(\rho_2^{e}-\rho_1^{o})(\rho_1^{o}-\rho_2^{o})}}\; , \;
\overline{\overline{\delta}}_s=1-\overline{\overline{\delta}}_r-\overline{\overline{\delta}}_l.
\end{array}
\end{equation}
in which:
\begin{equation}
\label{defk}
\fl
\eqalign{
k_{+}(x,y) =
{1 \over 2 x} \Bigl( & -x(1-q) + y (1-p) + p - q \cr
& + \sqrt{
\left (-x(1-q) + y (1-p) + p - q \right )^2 + 4 x y
\left (1-q\right )\left (1-p\right ) }
\Bigr) \, , \cr
}
\end{equation}
besides the following constraint on the reaction probabilities:
\begin{equation}
\label{defs2}
\fl
k_{+}(\beta,\delta)k_{+}(\alpha,\gamma)=\Bigl(\frac{p}{q}\Bigr)(1-p)(1-q).
\end{equation}
The equations of motion of the shock front (\ref{eqs}) indicate that the shock front hops to the left and right and also reflects from the boundaries of the lattice with above calculated probabilities. The constraint (\ref{defs2}) reminds us of the condition under which the steady-state of the PASEP can be written using the matrix product method and that the matrix representation is two-dimensional \cite{HP97}. \\
Since the dynamics of the shock front is simply similar to that of a single random walker, one can write the steady-state of the PASEP under the constraint (\ref{defs2}) as a linear superposition of the product measures (\ref{shocks}):
\begin{equation}
\label{linear sup}
\vert P^{\ast} \rangle = \frac{1}{Z}\sum_{k=1}^{2L+1}c_k \vert \mu_k \rangle
\end{equation}
in which one should have the steady-state condition as $T\vert P^{\ast} \rangle =\vert P^{\ast} \rangle$. Requiring this condition one finds:
\begin{equation}
\fl
\begin{array}{ll}
c_{1}=\frac{\delta_{r}^2}{\overline{\delta}_{r}(\delta_{r}+\delta_{l}\overline{\delta}_{r}-\delta_{r}\overline{\delta}_{l})},&
c_{2L+1}=\frac{\delta_{l}^2}{\overline{\overline{\delta}}_{l}((\delta_{r}+\delta_{l})\overline{\overline{\delta}}_{l}+
\delta_{l}\overline{\overline{\delta}}_{s})}(\frac{\delta_{r}}{\delta_{l}})^{2L},\\ \\
c_{2}=(\frac{\delta_{r}-\overline{\delta}_{r}(\delta_{r}+\overline{\delta}_{r}\delta_{l}-\overline{\delta}_{l}\delta_{r})-
\overline{\delta}_{l}\delta_{r}}{\delta_{r}+\delta_{l}\overline{\delta}_{r}-\delta_{r}\overline{\delta}_{l}})(\frac{\delta_{r}}{\delta_{l}}),&
c_{2L}=\frac{\delta_{l}(\delta_{s}\overline{\overline{\delta}}_{l}-(\delta_{l}-1)\overline{\overline{\delta}}_{s})}
{(\delta_{r}+\delta_{l})\overline{\overline{\delta}}_{l}+\delta_{l}\overline{\overline{\delta}}_{s}}(\frac{\delta_{r}}{\delta_{l}})^{2L-1},\\ \\
c_{2k+1}=(\frac{\delta_{r}}{\delta_{l}})^{2k} \; \;  k=1,\cdots,L-1,&c_{2k}=\delta_{s}(\frac{\delta_{r}}{\delta_{l}})^{2k-1} \;\; k=2,\cdots,L-1.
\end{array}
\end{equation}
The reader can easily calculate the normalization factor $Z$ in (\ref{linear sup}) using the normalization condition $Z=\sum_{k=1}^{2L+1}c_k$. \\
Let us now investigate the steady-state of the PASEP under the constraint (\ref{defs2}) using the matrix product method. Since the steady-state
of this system is unique, one should find the same stationary probability distribution vector as in (\ref{linear sup}). We adopt the notation used
in \cite{HP97} and write the matrix product steady-state of the system as follows:
\begin{equation}
\label{matrix prod}
\vert P^{\ast}\rangle= \frac{1}{Z} \langle \langle W \vert
\left[\left( \begin{array}{c}
E \\ D \end{array} \right) \otimes \left( \begin{array}{c}
\hat{E} \\ \hat{D} \end{array} \right)\right]^{\otimes L}
\vert V \rangle \rangle
\end{equation}
in which the operators $E$ and $D$ ($\hat{E}$ and $\hat{D}$ ) stand for the presence of an empty lattice site and a particle at odd (even) lattice sites, respectively.  The denominator $Z$ is a normalization factor.\\
It has been shown that for the PASEP on a finite lattice and in a discrete-time update, the operators ($\hat{E},\hat{D}$) and ($E,D$) besides the vectors $\vert V \rangle\rangle$ and $\langle \langle W \vert$ should satisfy the following quadratic algebra \cite{HP97} :
\begin{equation}
\fl
\begin{array}{l}
\label{BulkAlgebra}
[E,\hat{E}] = [D,\hat{D}] = 0 \; ,
\; (1-q)\hat{E} D+p\hat{D}E = E\hat{D} \; , \; q\hat{E}D+(1-p)\hat{D}E=D\hat{E}, \\
\langle\langle W \vert ((1-\alpha)E+\gamma D) = \langle \langle W \vert \hat{E} \; , \;
\langle\langle W \vert (\alpha E + (1-\gamma)D) = \langle \langle W \vert \hat{D} , \\
((1-\delta)\hat{E}+\beta \hat{D}) \vert V \rangle\rangle = E\vert V \rangle\rangle \; , \;
(\delta \hat{E}+(1-\beta)\hat{D}) \vert V \rangle \rangle= D\vert V \rangle\rangle.
\end{array}
\end{equation}
The quadratic algebra (\ref{BulkAlgebra}) results in $[E+D,\hat{E}+\hat{D}]=0$. It is easy now to see that the normalization factor $Z$ in (\ref{matrix prod}) can be also written in a grand canonical ensemble as:
\begin{equation}
Z=\langle \langle W \vert (E+D)^L(\hat{E}+\hat{D})^L \vert V\rangle \rangle.
\end{equation}
Note that in order to obtain the quadratic algebra (\ref{BulkAlgebra}), it is not needed the constraint (\ref{defs2}) to be hold; however, it can be readily verified that if the constraint (\ref{defs2}) is hold, then the following two-dimensional matrices and vectors satisfy the above mentioned quadratic algebra:
\begin{equation}
\begin{array}{l}
\label{BulkRep}
D = \left( \begin{array}{cccc}
\rho_{2}^{o}& & & 0 \\
\hat{d} & & & \frac{\delta_{r}}{\delta_{l}}\rho_{1}^{o}
\end{array} \right) \; , \;
E = \left( \begin{array}{cccc}
1-\rho_{2}^{o}& & & 0 \\
-\hat{d} & & & \frac{\delta_{r}}{\delta_{l}}(1-\rho_{1}^{o})
\end{array} \right) \; ,\; \\ \\
\hat{D} = \left( \begin{array}{cccc}
\rho_{2}^{e}  & & & 0 \\
d & & & \frac{\delta_{r}}{\delta_{l}}\rho_{1}^{e}
\end{array} \right) \; , \;
\hat{E} = \left( \begin{array}{cccc}
1-\rho_{2}^{e} & & & 0  \\
-d & & & \frac{\delta_{r}}{\delta_{l}}(1-\rho_{1}^{e})
\end{array} \right) \; , \; \\ \\
\langle\langle W \vert = (w_1, w_2) \; , \;
\vert V \rangle \rangle= \left( \begin{array}{c}
v_1 \\ v_2 \end{array} \right)
\end{array}
\end{equation}
provided that the matrix elements satisfy the following relations:
\begin{equation}
\label{defbound}
\fl
\frac{w_1}{w_2}=\frac{d-(1-\alpha-\gamma)\hat{d}}{\alpha-\rho_{2}^{e}+(1-\alpha-\gamma)\rho_{2}^{o}}(\frac{\delta_l}{\delta_r}) \; , \\
\frac{v_1}{v_2}=\frac{\delta-\rho_{1}^{o}+(1-\beta-\delta)\rho_{1}^{e}}{\hat{d}-(1-\beta-\delta)d}(\frac{\delta_r}{\delta_l}).
\end{equation}
Note that in (\ref{BulkRep}) and (\ref{defbound}) the densities of the particles on the sides on the shock front and also the shock front hopping probabilities are given in (\ref{defs1}). The matrix representation structure given in (\ref{BulkRep}) was first introduced in \cite{JM09} for the Totally Asymmetric Simple Exclusion Process (TASEP) with open boundaries in sublattice-parallel updating scheme. In the TASEP the particles enter the system only from the left boundary (the first lattice site). The particles perform a totally biased diffusion only toward the right boundary and leave the system only from there (the last lattice site). As can be seen, the same matrix structure is still valid in order to explain the steady-state of the PASEP when it can be written in terms of a linear superposition of shocks introduced in (\ref{shocks}). In comparison to the TASEP, the shock front in the PASEP reflects from the boundaries with more complicated probabilities.\\
As it was stated in \cite{JM07}, at least for the one-dimensional systems with nearest neighbor interactions and in continuous-time updating scheme, the matrix product steady-state can be constructed using two-dimensional matrices with a generic structure if it is made up of a linear combination of single product shock measures with random walk dynamics. For the same systems, but in discrete-time updating scheme, it seems that the two-dimensional matrix representation first proposed in \cite{JM09} is sufficient in order to construct the steady-state. Considering the example studied here i.e. the PASEP on a finite lattice with open boundaries besides the example studied in \cite{JM09}, it seems that the same statement can be applied to the above mentioned systems in discrete-time updating scheme regardless of the microscopic reaction rules. Using the two-dimensional matrix representation presented in (\ref{BulkRep}) one can calculate any physical quantity such as the current and the density profile of the particles. Clearly one finds the same results obtained in \cite{HP97} since this representation can be obtained from the one used in \cite{HP97} by applying a similarity transformation. The main difference is that the matrix representation used here is much simpler than the one in \cite{ER96,HP97} and that it clearly reveals the shock characteristics of the system.
\section*{References}

\end{document}